\documentclass[10pt]{article}

\usepackage{amsmath}
\usepackage{amssymb}
\usepackage{graphics}
\usepackage{rotating}
\usepackage{cite}
\usepackage{color}
\usepackage{fancybox}
\usepackage{pstricks}
\usepackage{xcolor}
\usepackage{shadowtext}
\usepackage{multicol}

\def\0{\mbox{\tiny $0$}}
\def\1{\mbox{\tiny $1$}}
\def\2{\mbox{\tiny $2$}}
\def\3{\mbox{\tiny $3$}}
\def\4{\mbox{\tiny $4$}}
\def\5{\mbox{\tiny $5$}}
\def\6{\mbox{\tiny $6$}}
\def\7{\mbox{\tiny $7$}}
\def\8{\mbox{\tiny $8$}}
\def\9{\mbox{\tiny $9$}}
\definecolor{navy}{rgb}{0,0,.6}
\definecolor{jour}{rgb}{0,0.6,.4}
\definecolor{jbul}{rgb}{0.7,0.,.4}
\begin{document}
%
\thispagestyle{empty}
\setcounter{page}{0}

\begin{center}
\shadowrgb{0.8,0.8,1}
\shadowoffset{4pt}
\shadowtext{
\color{navy}
\fontsize{18}{18}\selectfont
\bf THE USE OF THE STATIONARY PHASE METHOD}\\
\shadowtext{
\color{navy}
\fontsize{18}{18}\selectfont
\bf AS A MATHEMATICAL TOOL TO DETERMINE}\\
\shadowtext{
\color{navy}
\fontsize{18}{18}\selectfont
\bf THE PATH OF OPTICAL BEAMS}
\end{center}

\vspace*{1cm}

\begin{center}
\shadowrgb{0.8, .8, 1}
\shadowoffset{2.5pt}
\shadowtext{\color{jbul}
\fontsize{13}{13}\selectfont
$\boldsymbol{\bullet}$}
\shadowtext{\color{jour}
\fontsize{15.5}{15.5}\selectfont
\bf American Journal of Physics 83, 249-255 (2015)}
\shadowtext{\color{jbul}
\fontsize{13}{13}\selectfont
$\boldsymbol{\bullet}$}
\end{center}

\vspace*{1cm}

\begin{center}
\begin{tabular}{cc}
\begin{minipage}[t]{0,5\textwidth}
{\bf Abstract}.
We use the stationary phase method to determine the path of optical beams which propagate through a dielectric block. In the presence of partial internal reflection, we recover the geometrical result obtained by using the Snell law. For total internal reflection, the stationary phase method overreaches the Snell law predicting the Goos-H\"anchen shift.
\end{minipage}
& \begin{minipage}[t]{0,5\textwidth}
{\bf Silv\^ania A. Carvalho}\\
Department of Applied Mathematics\\
State University of Campinas (Brazil)\\
{\color{navy}{{\bf  silalves@ime.unicamp.br}}}
\hrule
\vspace*{0.2cm}
{\bf Stefano De Leo}\\
Department of Applied Mathematics\\
State University of Campinas (Brazil)\\
{\color{navy}{{\bf  deleo@ime.unicamp.br}}}
\end{minipage}
\end{tabular}
\end{center}

\vspace*{2cm}

\begin{center}
\shadowrgb{0.8, 0.8, 1}
\shadowoffset{2pt}
\shadowtext{\color{navy}
{\bf
\begin{tabular}{ll}
I. & INTRODUCTION \\
II. & THE OPTICAL PATH VIA SNELL'S LAW \\
III. & MAXWELL EQUATIONS AND TRANSMISSION COEFFICIENT \\
IV.  & THE OPTICAL PATH VIA THE STATIONARY PHASE METHOD \\
V. & CONCLUSIONS AND OUTLOOKS\\
& \\
& \,[\,11 pages, 2 figures\,]
\end{tabular}
}}
\end{center}

\vspace*{7cm}

\begin{flushright}
\shadowrgb{.8, .8, 1}
\shadowoffset{2pt}
\shadowtext{\color{jbul}
\fontsize{11}{11}\selectfont
$\boldsymbol{\bullet}$}
\hspace*{-.2cm}
\shadowtext{
\color{jour}
\fontsize{15.5}{15.5}\selectfont
$\boldsymbol{\Sigma\hspace*{0.06cm}\delta\hspace*{0.035cm}\Lambda}$}
\hspace*{-.35cm}
\shadowtext{\color{jbul}
\fontsize{11}{11}\selectfont
$\boldsymbol{\bullet}$}
\end{flushright}


\newpage


\section*{\normalsize I. INTRODUCTION}

Optical beams propagating through dielectrics with dimensions greater than the
wavelength of light can be described by rays obeying a set of geometrical
rules.\cite{Wolf1999,Saleh2007,Snell2012} In this limit, Snell's law is used to
determine the relationship between the incidence and refraction angles and,
consequently, the paths of optical
beams.\cite{Snell1948,Snell1951,Snell1958,Snell1980,Snell2005,Snell2007} In this
paper, we discuss beam propagation through the dielectric block illustrated in
Fig.~1 and, as the first step in our analysis (Sec.~II), we calculate its
geometrical path using Snell's law.

The analogy between optics\cite{Wolf1999,Saleh2007} and quantum
mechanics\cite{Shiff1955,Cohen1977} has been a matter of discussion in recent
works.\cite{RW2009,RW2013} The possibility of linking Maxwell's equations for photon
propagation in the presence of dielectric blocks with the quantum-mechanical
equations for the propa\-gation of electrons in the presence of potential
steps\cite{QM1957,QM2011,QM2012} allows one to determine, in a simple and intuitive
way,  the reflection and transmission coefficients at the dielectric block
interfaces.\cite{Use2008,Use2011a,Use2011b,Use2013}  As the second step in our
analysis (Sec.~III), we obtain the reflection and transmission coefficients for
$s$-polarized waves transmitted through the dielectric system of Fig.~1.

Once we obtain the transmission coefficient at the exit interface, we use the
stationary phase method (SPM) to calculate the optical path by imposing the
cancellation (in the oscillatory electric field integral) of sinusoids with rapidly
varying phase.\cite{SPM1955,SPM1973,SPM1975} The calculation of the position of the
maximum of the outgoing beam at the exit of the dielectric block, based on the SPM
(Sec.~IV), represents an alternative way to obtain the optical path, one that does
not require a geometrical analysis. The SPM analysis only requires that we cancel
the derivative of the outgoing beam phase.  However, the use of the SPM as a
mathematical tool to obtain the path of optical beams propagating into dielectric
blocks is not simply a matter of taste. For total internal
reflection,\cite{TIR1981,TIR2005} the SPM also predicts the Goos-H\"anchen (GH)
shift\cite{GHS1947,GHS1988,GHS2012,GHS2013} (see Fig.~2). This shows the importance
of the SPM not only to recover Snell's law but also to obtain a typical
quantum-mechanical effect. The SPM, illustrated in this paper for calculating the
path of optical beams, is a mathematical tool easily extended to other fields of
physics  in which wave packets play an important role.

Finally, after discussing our conclusions, we extend our results to  $p$-polarized waves,
suggest how to amplify the GH shift by building a band of dielectric blocks, and propose further
theoretical investigations.

\section*{\normalsize II. THE OPTICAL PATH VIA SNELL'S LAW}

Let us consider an incoming gaussian beam, with beam waist size $w_{0}$ and wavenumber $k$, which moves along the $z$-direction:\cite{Wolf1999,Saleh2007}
\begin{eqnarray}
E_{_{\rm in}}(\textbf{r}) &=& E_{0}\,e^{ikz} \frac{w_{0}^{2}}{w_{0}^{2}+2i z/k} \exp \left(-\frac{x^{2} +
y^{2}}{w_{0}^{2}+2i z/k }\right),
\label{eq:Enew}
\end{eqnarray}
where
\begin{equation}
\label{eqr}
\textbf{r} = x\,\textbf{e}_x+y\,\textbf{e}_y+z\,\textbf{e}_z.
\end{equation}
The plane of incidence is chosen to  be the $yz$-plane. The normals to the left/right and up/dwon sides of the dielectric block are
respectively oriented along the direction of the unit vectors $\textbf{e}_{\tilde{z}}$ and $\textbf{e}_{z_*}$, see Fig.~1\,(a). The unit vector $\textbf{e}_{\tilde{z}}$  forms an angle $\theta$ with $\textbf{e}_{z}$,
 \begin{eqnarray}
\textbf{e}_{\tilde{y}} &=&  \phantom{-}\textbf{e}_y \cos\theta + \textbf{e}_z \sin\theta, \\
\textbf{e}_{\tilde{z}} &=&  -\,\textbf{e}_y \sin\theta + \textbf{e}_z \cos\theta,
\end{eqnarray}
and $\textbf{e}_{z_*}$  an angle $\pi/4$ with $\textbf{e}_{\tilde{z}}$,
 \begin{eqnarray}
\textbf{e}_{y_*} &=&  (\phantom{-\,}\textbf{e}_{\tilde{y}} + \textbf{e}_{\tilde{z}} )/\sqrt{2}, \\
\textbf{e}_{z_*} &=&  (-\,\textbf{e}_{\tilde{y}}  + \textbf{e}_{\tilde{z}})/\sqrt{2}.
\end{eqnarray}
In the new coordinates systems, the components of the position vector $\textbf{r}$
are then related by
\begin{eqnarray}
\left(\begin{array}{c} y_*\\ z_* \end{array} \right) &=& \frac{1}{\sqrt{2}}
\left(\begin{array}{rr} 1 & 1 \\ -1 & 1
\end{array} \right)\left(\begin{array}{c} \tilde{y}\\ \tilde{z} \end{array} \right)=
\frac{1}{\sqrt{2}}
\left(\begin{array}{rr} 1 & 1 \\ -1 & 1
\end{array} \right)
\left(\begin{array}{rr} \cos \theta & \sin\theta \\ -\sin \theta & \cos\theta
\end{array} \right)\left(\begin{array}{c} y\\ z \end{array} \right)
\nonumber \\
 & \equiv & \textbf{R}\left(\frac{\pi}{4}+\theta\right)
\left(\begin{array}{c} y\\ z \end{array} \right)
\end{eqnarray}
where $\textbf{R}$ represents the rotation matrix. Observe that $x=\tilde{x}=x_*$.  The new coordinates systems, in this Section, will be used to calculate the geometrical path by the Snell law and in the following section to determine the reflection and transmission coefficients of the optical beam.

In order to determine the optical path by the Snell law,  we first demonstrate that the outgoing beam is {\em parallel} to the incoming one, then we calculate the exit point $P_{_{\rm right}}$ and finally we obtain the distance $d$ between the incoming and outgoing beams, see Fig.\,1(a).

When an optical beam falls onto a boundary between two homogeneous media with
different refractive indices, it is split into two beams. The refracted
(transmitted) beam propagates into the second medium and the reflected beam
propagates back into the first medium. Snell's law states that the ratio of the
sines of the angles of incidence and refraction  is equal to the
reciprocal of the ratio of the refraction indices of the media in which the beams
propagate.  Thus, for the left air/dielectric boundary we have
\begin{equation}
\sin\theta = n \sin\psi\,,
\end{equation}
where $n$ is the index of refraction for the dielectric material, $\theta$ and $\psi$  are the angles that the incoming and refracted beams form with $\textbf{e}_{\tilde{z}}$.  In this case, the second medium is optically denser than the first and consequently the refracted angle is a real quantity for all incident angles.  The beam propagates into the dielectric block at an angle $\psi$ with the $\tilde{z}$-axis  and an angle $\pi/4+\psi$ with the $z_*$-axis, see Fig.~1\,(a).   Due to the fact that the triangles $AP_{_{\rm left}}P_{_{\rm down}}$ and $CP_{_{\rm right}}P_{_{\rm up}}$ are similar,
\begin{equation}
A\widehat{P}_{_{\rm left}}P_{_{\rm down}}=C\widehat{P}_{_{\rm right}}P_{_{\rm up}}=\frac{\pi}{4}-\psi \equiv \alpha.
\end{equation}
 Consequently, the optical beam forms an angle $\psi$ with the normal to the right side of the dielectric block. By using the Snell law, we then find that the outgoing beam forms an angle $\theta$ with $\textbf{e}_{\tilde{z}}$. This implies that the outgoing beam is parallel to the incoming one.

 Let us not determine the point in which the outgoing beam  leaves the block, i.e. $P_{_{\rm right}}$. To do it,  the better coordinates system  choice is represented by the $\{\textbf{e}_{y_*}, \textbf{e}_{z_*}\}$ system.  Without loss of generality, we can  choose as origin  of the coordinates system the incident point on the left (air/dielectric) interface:
\begin{equation}
P_{_{\rm left}}=\{\,0\,,\,0\,\}.
\end{equation}
By simple geometrical considerations, we immediately get
\begin{equation}
P_{_{\rm down}}= \overline{P_{_{\rm left}}A}\,\sin \frac{\pi}{4}\,\left\{\,\frac{1}{\tan \alpha}\,,\,1\,\right\} = \frac{a}{\sqrt{2}}\,\left\{\, \tan\left(\frac{\pi}{4} + \psi \right) \,,\, 1  \,\right\}
\end{equation}
and
\begin{equation}
P_{_{\rm up}} = P_{_{\rm down}}+\,\overline{AB}\,\sin \frac{\pi}{4}\,\left\{\,\frac{1}{\tan \alpha}\,,\,-\,1\,\right\} = \frac{1}{\sqrt{2}}\,\left\{\, (a+b)\tan\left(\frac{\pi}{4} + \psi \right) \,,\, a-b \,\right\}.
\end{equation}
To obtain the coordinates of  $P_{_{\rm right}}$ we need to calculate the intersection between the following straight lines $r\left(P_{_{\rm up}}P_{_{\rm right}} \right)$ and   $r\left( P_{_{\rm right}}C\right)$,
\[  r\left(P_{_{\rm up}}P_{_{\rm right}}  \right) \,\bigcap \,r\left( P_{_{\rm right}}C\right).  \]
In the $\{\textbf{e}_{y_*}, \textbf{e}_{z_*}\}$ system, the  straight lines are represented by
\begin{eqnarray}
 r\left(P_{_{\rm up}}P_{_{\rm right}} \right)& :\,\, & z_{*} - \frac{a - b}{\sqrt{2}} =  \tan\left(\frac{\pi}{4} - \psi \right)\, \left[\,y_{*} - \frac{b + a}{\sqrt{2}}\tan
\left(\frac{\pi}{4} + \psi \right) \right],\\
 r\left(P_{_{\rm right}}C \right)& :\,\, & z_{*} - \frac{a - b}{\sqrt{2}} = -\, \left[\,y_{*} - \left(\frac{b - a}{\sqrt{2}} + c\,\sqrt{2} \right) \right].
\end{eqnarray}
After simple algebraic manipulations, the previous equation can be simplified to
\begin{eqnarray}
 r\left(P_{_{\rm up}}P_{_{\rm right}} \right)& :\,\, & z_{*} =  \tan\left(\frac{\pi}{4} - \psi \right)\,y_{*} - b\,\sqrt{2}, \label{s1}\\
 r\left(P_{_{\rm right}}C \right)& :\,\, & z_{*} = -\, y_{*} + c\,\sqrt{2}\label{s2}.
\end{eqnarray}
Finally, $P_{_{\rm right}}$ can be determined by solving the system
(\ref{s1}) and (\ref{s2}),
\begin{equation}
P_{_{\rm right}} = \left\{\,\frac{\left(\,b + c\,\right)\,\sqrt{2}}{1 + \tan
\left(\frac{\pi}{4} - \psi \right)}\,,\, \frac{\left[\,c \tan \left(\frac{\pi}{4} -
\psi \right)-b\,\right]\,\sqrt{2}}{1 + \tan \left(\frac{\pi}{4} - \psi
\right)}\,\right\}.
\label{eq:pright}
\end{equation}
Once known $P_{_{\rm right}}$, observing that
\begin{equation}
 r\left(SP_{_{\rm left}} \right):\,\, z_* = \tan \left(\frac{\pi}{4} - \theta
\right)\,y_*
\end{equation}
we can immediately calculate the shift between the incoming and outgoing beams,
\begin{eqnarray}
\label{dfor}
d\left[P_{_{\rm right}}, r\left(SP_{_{\rm left}} \right)\right] & = &
\left[\,\tan\left(\frac{\pi}{4} - \theta \right)\,y_*\left(P_{_{\rm right}}\right)
- z_*\left(P_{_{\rm right}}\right)\,\right]\,\mbox{\Large
$/$}\sqrt{1+\tan^{2}\left(\frac{\pi}{4} - \theta \right)} \nonumber \\
   & = &
\left[\,b+(b+c)\,\frac{\sin\theta}{\sqrt{n^2-\sin^{2}\theta}}\,\right]\,\cos\theta\,
-\,c\,  \sin\theta\,.
\label{eq:eqdsl0}
\end{eqnarray}
Before concluding this section, we note that, depending on the incidence angle
$\theta$,
the internal reflections can be partial or total. Let us explain this in detail by
calculating the critical angle which characterizes the limit angle between partial
and total reflection.
As observed at the begin of this section, at the first air/dielectric interface the
second medium ($n>1$) is optically denser than the first one ($n=1$) and we always
find
a refracted beam which moves into the dielectric block forming  a
real angle $\psi=\arcsin(\,\sin\theta/n)$  with the $\tilde{z}$-axis. At the second interface, $P_{_{\rm
down}}$, we have {\em total internal reflection} when
\begin{equation}
\label{con1}
n\,\sin\left(\frac{\pi}{4}+\psi\right)>1.
\end{equation}
It is important to note here that for $\psi>\pi/4$ the refracted beam cannot reach
the second interface. This represents  an additional constraint to be considered in our
discussion. By adding this constraint to Eq.(\ref{con1}),  we obtain the condition for total
internal reflection,
\begin{equation}
\label{cond}
\arcsin\left(\frac{1}{n}\right) - \frac{\pi}{4} < \psi < \frac{\pi}{4}.
\end{equation}
In terms of the incidence angle $\theta$, the previous condition becomes
\begin{equation}
\arcsin\left(\,\frac{1-
\sqrt{n^2-1}}{\sqrt{2}}\,\right)<\theta<\arcsin\left(\frac{n}{\sqrt{2}}
\right).
\end{equation}
In Fig.\,2(a), we plot the critical angle
\begin{equation}
\theta_c  = \arcsin\left(\,\frac{1-
\sqrt{n^2-1}}{\sqrt{2}}\,\right)
\end{equation}
 as a function of the refractive index $n$. This curve separates the partial and total reflection zones. We conclude this section, by observing that for
\begin{equation*}
\frac{1- \sqrt{n^2-1}}{\sqrt{2}} \leq -1,
\end{equation*}
which implies
\begin{equation}
\label{eq:cri1}
n\geq \sqrt{4+2\sqrt{2}},
\end{equation}
we {\em always} find total internal reflection, see Fig.\,2(a).

\section*{\normalsize III. MAXWELL EQUATIONS AND TRANSMISSION COEFFICIENT}

In this section, by using the Maxwell equations\cite{Wolf1999,Saleh2007}
we calculate the transmission and reflection coefficient at each interface. The
phase of the outgoing beam will be then used to calculate by the stationary phase
method\cite{SPM1955,SPM1973,SPM1975}  the path of the optical beam.

The plane wave solution of
\begin{equation}
\left[\, \partial_{xx} + \partial_{yy} + \partial_{zz} -
\frac{\partial_{tt}}{c^2}\right]\, E (\textbf{r},t) = 0
\label{eq:eqME}
\end{equation}
is given by
\begin{equation}
\exp\left[\,i\, (\,k_{x} x + k_{y} y + k_{z} z -\omega\, t\,)\,\right]\,,
\end{equation}
providing that
\begin{equation*}k=\sqrt{k_x^2+k_y^2+k_z^2} =\omega/c \,. \end{equation*}
Obviously a convolution of these plane waves is also a solution of Eq.(\ref{eq:eqME}). Many lasers emit beams that approximate a gaussian profile\cite{WP2008,Use2011a,Use2011b}
\begin{equation}
\label{ginc}
E (\textbf{r},t) =
E_{0}\,\frac{w_{0}^{{2}}}{4\,\pi}\int\mbox{d}k_x\mbox{d}k_y\,\exp\left[\,-\,
\left(\,k_x^2+k_y^2\,\right)\,
w_{0}^{2}/\,4\,\right]\,\exp[\,i\,(\,\boldsymbol{k}\cdot\textbf{r}-\omega\,t\,)\,]\,,
\end{equation}
where  $w_{0}$ is the beam waist size. Observe that for $\sqrt{k_x^{^{2}}+k_y^{^{2}}}\ll k$ (paraxial approximation) the previous integral can be calculated analytically and the integration leads to Eq.(\ref{eq:Enew}).

Due to the fact that the first air/dielectric discontinuity is along the $\tilde{z}$-axis, it is
convenient to rewrite the Maxwell equations  by using the coordinate system $(x,\tilde{y},\tilde{z})$,
\begin{equation}
\left[\, \partial_{xx} + \partial_{\tilde{y}\tilde{y}} +
\partial_{\tilde{z}\tilde{z}} - n^{2}(\tilde{z})\,\frac{\partial_{tt}}{c^2}\right]\,
E_{_{\rm left}} (\textbf{r},t) =
0\,,\,\mbox{with}\,n(\tilde{z})=\left\{\begin{array}{l}
1\,\mbox{for}\,
\tilde{z}<0\,,\\
n\,\mbox{for}\,
\tilde{z}>0\,.
\end{array}
\right.
\label{MEleft1}
\end{equation}
 The plane wave solution is now given by
\begin{equation}
\label{eqleft}
\exp\left[\,i\, (\,k_{x} x + k_{\tilde{y}}\, \tilde{y} -\omega\, t\,)\,\right]
\,\times \,\begin{cases}
\exp\left[\,i\,k_{\tilde{z}}\, \tilde{z}\,\right] + R_{_{\rm
left}}\,\exp\left[\,-\,i\,k_{\tilde{z}}\, \tilde{z}\,\right] &
\mbox{for}\,\tilde{z}<0 \,,\\
T_{_{\rm left}}\,\exp\left[\,i\,q_{\tilde{z}}\, \tilde{z}\,\right] &
\mbox{for}\,\tilde{z}>0 \,,
\end{cases}
\end{equation}
where
 \begin{equation*}
 \left(\,\begin{array}{c} k_{\tilde{y}}\\ k_{\tilde{z}} \end{array} \right) =
\,{\bf R}[\,\theta\,]\,
\left(\,\begin{array}{c} k_y\\ k_z \end{array} \right)
\hspace{0.6cm}\mbox{and}\hspace{0.6cm}
\left\{\,q_{\tilde{y}}\, ,\,q_{\tilde{z}}\,\right\}=\left\{\,k_{\tilde{y}}\,,\,\sqrt{n^2k^2 - k_x^2- k_{\tilde{y}}^2}\,\right\}\,.
 \end{equation*}
 Observe that the $\tilde{y}$-component of the wave number does not change because
the discontinuity is along the $\tilde{z}$-axis\cite{Cons1951}. By matching the
function (\ref{eqleft}) imposing the continuity of $E_{\rm left}$ and
$\partial_{\tilde{z}}\,E_{\rm left}$ at the point where the refractive index is
discontinuous, i.e. $\tilde{z}=0$, we find
\begin{equation}
R_{_{\rm left}} = \frac{k_{\tilde{z}} \,-\, q_{\tilde{z}}}{k_{\tilde{z}}\, +\,
q_{\tilde{z}}} \hspace{0.9cm} \mbox{and} \hspace{0.9cm} T_{_{\rm left}} = \frac{2\,
k_{\tilde{z}}}{k_{\tilde{z}} \,+\, q_{\tilde{z}}}\,.
\end{equation}
At the second interface it is convenient to use the coordinate  system
$(x,y_*,z_*)$ and solve
the Maxwell equation
\begin{equation}
\left[\, \partial_{xx} + \partial_{y_*y_*} + \partial_{z_*z_*} -
n^{2}(z_*)\,\frac{\partial_{tt}}{c^2}\right]\, E_{_{\rm down}} (\textbf{r},t) =
0\,,\,\mbox{with}\,n(z_*)=\left\{\begin{array}{l}
n\,\mbox{for}\,
z_*<a/\sqrt{2}\,,\\
1\,\mbox{for}\,
z_*>a/\sqrt{2}\,.
\end{array}
\right.
\label{MEleft2}
\end{equation}
The plane wave solution is
\begin{equation}
\label{eqdown}
\exp\left[\,i\, (\,k_{x} x + q_{y_*}\, y_* -\omega\, t\,)\,\right] \times \begin{cases}
\exp\left[\,i\,q_{z_*}\, z_*\,\right] + R_{_{\rm down}}\,\exp\left[\,-\,i\,q_{z_*}\,
z_* \,\right] &  \mbox{for}\,z_*<a/\sqrt{2}\,,\\
T_{_{\rm down}}\,\exp\left[\,i\,k_{z_*}\, z_*\,\right] & \mbox{for}\,z_*>a/\sqrt{2}\,,
\end{cases}
\end{equation}
where
\begin{equation*}
 \left(\,\begin{array}{c} q_{y_*}\\ q_{z_*} \end{array} \right) =  \,{\bf
R}[\,\mbox{$\frac{\pi}{4}$}\,]\,
\left(\,\begin{array}{c} k_{\tilde{y}}\\ q_{\tilde{z}} \end{array} \right)
\hspace{0.6cm}\mbox{and}\hspace{0.6cm}
\left\{\,k_{y_*}\,,\,k_{z_*}\,\right\} =\left\{\,q_{y_*}\,, \sqrt{k^2 - k_x^2- q_{y_*}^2}\,\right\}\,.
 \end{equation*}
As happens for the first interface, the
$y_*$-component of the wave number is not modified  because the discontinuity of the
second interface is along the $z_*$-axis. By matching the function (\ref{eqdown}) at
  $\tilde{z}=a/\sqrt{2}$, we find
\begin{equation}
R_{_{\rm down}} = \frac{q_{z_{*}} \,-\, k_{z_{*}}}{q_{z_{*}} \,+\, k_{z_{*}}}
\,\exp\left[\,2 \,i \,q_{z_{*}}\, \frac{a}{\sqrt{2}}\,\right] \hspace{0.6cm}
\mbox{and} \hspace{0.6cm} T_{_{\rm down}} = \frac{2\, q_{z_{*}}}{q_{z_{*}}\, +\,
k_{z_{*}}} \exp\left[i \,( q_{z_{*}} -k_{z_{*}} )\, \frac{a}{\sqrt{2}}\right]\,.
\end{equation}
For the up interface, we can use the result obtained for the down interface. By
replacing
\begin{equation*}(\,k_{z_*}\,,\,q_{z_*}\,)\,\to\,-\,(\,k_{z_*}\,,\,q_{z_*}\,)\hspace{0.6cm}\mbox{and}
\hspace{0.6cm}\frac{a}{\sqrt{2}}\,\to\,\frac{a-b}{\sqrt{2}}\,,\end{equation*}
we find the reflection and transmission coefficients for the up interface,
\begin{equation}
R_{_{\rm up}} = \frac{q_{z_{*}} \,-\, k_{z_{*}}}{q_{z_{*}} \,+\, k_{z_{*}}}
\,\exp\left[\,2 \,i \,q_{z_{*}}\, \frac{b-a}{\sqrt{2}}\,\right] \hspace{0.6cm}
\mbox{and} \hspace{0.6cm} T_{_{\rm up}} = \frac{2\, q_{z_{*}}}{q_{z_{*}}\, +\,
k_{z_{*}}} \exp\left[i \,( q_{z_{*}} -k_{z_{*}} )\, \frac{b-a}{\sqrt{2}}\right]\,.
\end{equation}
In a similar way, the reflection and transmission coefficients for the right
interface can be directly obtained from the coefficients calculated for the left
interface. By  replacing
\begin{equation*} k_{\tilde{z}}\,\leftrightarrow\,q_{\tilde{z}}\end{equation*}
and observing that  the discontinuity is now located at $\tilde{z}=c$, we obtain
\begin{equation}
R_{_{\rm right}} = \frac{q_{\tilde{z}} \,-\, k_{\tilde{z}}}{q_{\tilde{z}} \,+\,
k_{\tilde{z}}} \exp\left[\,2 \,i \,q_{\tilde{z}}\, c\, \right] \hspace{0.9cm}
\mbox{and} \hspace{0.9cm} T_{_{\rm right}} = \frac{2
\,q_{\tilde{z}}}{q_{\tilde{z}}\, +\, k_{\tilde{z}}} \exp\left[\,i \,( q_{\tilde{z}}
-k_{\tilde{z}} )\, c\, \right]\,.
\label{eq:RTright}
\end{equation}
Finally, the outgoing transmission coefficient is
\begin{equation}
T_{\rm out} =  T_{_{\rm left}}\, R_{_{\rm down}}\,R_{_{\rm up}}\, T_{_{\rm right}}
 = \frac{4 \,k_{\tilde{z}} \,q_{\tilde{z}}}{(q_{\tilde{z}} \,+\, k_{\tilde{z}})^2}
\left(\frac{q_{z_{*}} \,-\, k_{z_{*}}}{q_{z_{*}} \,+ \, k_{z_{*}}}\right)^{2}
\exp[\,i \,q_{z_{*}}\,b\,\sqrt{2}\, + i \,(\,q_{\tilde{z}} \,-\, k_{\tilde{z}}\,)\,
c\,]\,.
\end{equation}
The spatial phases of the optical beam in the different regions are
\begin{equation*}
\begin{array}{rcl}
\mbox{in}\,)\, & \hspace*{.5cm} & k_x\, x \,+ \,k_y\,y\,+\,k_z\,z \,=\, k_x\, x \,+
\,k_{\tilde{y}}\,\tilde{y}\,+\,k_{\tilde{z}}\,\tilde{z}\,,\\
\mbox{left$\,\to\,$down}\,)\, &  & k_x\, x \,+
\,k_{\tilde{y}}\,\tilde{y}\,+\,q_{\tilde{z}}\,\tilde{z} \,=\,k_x\, x \,+
\,q_{y_*}\,y_*\,+\,q_{z_*}\,z_*
\,,\\
\mbox{down$\,\to\,$up}\,)\, &  &
k_x\, x \,+ \,q_{y_*}\,y_*\,-\,q_{z_*}\,z_*
\,,\\
\mbox{up$\,\to\,$right}\,)\, &  &
k_x\, x \,+ \,q_{y_*}\,y_*\,+\,q_{z_*}\,z_* \,=\, k_x\, x \,+
\,k_{\tilde{y}}\,\tilde{y}\,+\,q_{\tilde{z}}\,\tilde{z}
\,,\\
\mbox{out}\,)\, & \hspace*{.5cm} &  k_x\, x \,+
\,k_{\tilde{y}}\,\tilde{y}\,+\,k_{\tilde{z}}\,\tilde{z}\,=\,
k_x\, x \,+ \,k_y\,y\,+\,k_z\,z\,.
\end{array}
\end{equation*}
The outgoing beam, which as expected  is parallel to the
incoming one, is then given by\cite{Use2011a,Use2011b}
\begin{eqnarray}
\label{gout}
E_{\rm out} (\textbf{r},t) &=& E_{0} \,
\frac{w_{0}^{{2}}}{4\,\pi}\int\mbox{d}k_x\mbox{d}k_y\,
\frac{4 \,k_{\tilde{z}} \,q_{\tilde{z}}}{(q_{\tilde{z}} \,+\, k_{\tilde{z}})^2}
\left(\frac{q_{z_{*}} \,-\, k_{z_{*}}}{q_{z_{*}} \,+ \, k_{z_{*}}}\right)^{2}
\exp\left[\,-\,
\left(\,k_x^2+k_y^2\,\right)\, w_{0}^{2}/\,4\,\right]\,\times \nonumber \\\
& & \hspace*{2.7cm}\exp\{\,i\,[\,q_{z_{*}}\,b\,\sqrt{2}\, + \,(\,q_{\tilde{z}} \,-\,
k_{\tilde{z}}\,)\, c \,+\, \boldsymbol{k}\cdot\textbf{r} \,-\,
\omega\,t\,]\,\}\,.
\end{eqnarray}
In the next section, by using the SPM, we calculate the position of the maximum of
the outgoing beam and consequently the position of the optical beam at the exit of
our dielectric system. The calculation based on the SPM thus represents an
alternative way (with respect to the standard Snell law) to obtain the optical path.
More important, the SPM calculation also allows one to obtain  the GH-shift.

\section*{\normalsize IV. THE OPTICAL PATH VIA THE STATIONARY PHASE METHOD}

The SPM is a basic principle of asymptotic analysis which applies to oscillatory
integrals. The main idea of the SPM relies on the cancelation of sinusoids with
rapidly varying phase. To illustrate this principle let us consider the incoming
beam given in Eq.\,(\ref{ginc}). In order to maximize the integral we impose that
\begin{equation*}\left[\,\frac{\partial\,}{\partial k_x}\,
\left(\,\boldsymbol{k}\cdot
\textbf{r}\,-\,\omega\,t\,\right)\,\right]_{_{(0,0)}}=\,\left[\,
\frac{\partial\,}{\partial k_y}\, \left(\,\boldsymbol{k}\cdot
\textbf{r}\,-\,\omega\,t\,\right)\,\right]_{_{(0,0)}}=\,0\,. \end{equation*}
The derivatives have to be calculated at the maximum value of the convolution
function.  For the incoming optical beam of Eq.(\ref{ginc}),  the convolution
function is  a gaussian distribution, consequently, the maximum of the incoming beam
is located at
\begin{equation*}x=y=0\,.\end{equation*}
This result, obtained without any integration, is confirmed by Eq.\,(\ref{eq:Enew}).

\subsection*{\small IV.A PARTIAL INTERNAL REFLECTION}

As discussed in section II, for $\theta<\theta_c$, we get partial internal
reflection. In this case, the outgoing optical beam has, with respect to the
incoming beam, an additional phase given by
\begin{equation}
\label{phase}
\phi =\, q_{z_{*}}\,b\,\sqrt{2}\, + \,(\,q_{\tilde{z}} \,-\, k_{\tilde{z}}\,)\, c\,.
\end{equation}
By using the SPM, in order to maximize the outgoing integral, we have now to impose
the following constraints
\begin{equation*}\left[\,\frac{\partial\,}{\partial k_x}\,
\left(\,\phi\,+\,\boldsymbol{k}\cdot
\textbf{r}\,-\,\omega\,t\,\right)\,\right]_{_{(0,0)}}\hspace*{-0.2cm}=\,\left[\,
\frac{\partial\,}{\partial k_y}\, \left(\,\phi\,+\,\boldsymbol{k}\cdot
\textbf{r}\,-\,\omega\,t\,\right)\,\right]_{_{(0,0)}}\hspace*{-0.2cm}=\,0\,.
\end{equation*}
Observing that
\begin{eqnarray}
\label{eqder}
\frac{\partial q_{\tilde{z}}}{\partial k_{x,y}}& =& -\, \frac{
k_{\tilde{y}}}{q_{\tilde{z}}}\, \frac{\partial k7_{\tilde{y}}}{\partial k_{x,y}}  =
 -\, \frac{ k_{\tilde{y}}}{q_{\tilde{z}}}\,\left[\,\cos\theta\,  \frac{\partial
k_y}{\partial k_{x,y}} + \sin\theta\,\frac{\partial k_z}{\partial k_{x,y}} \,\right]
\,,\nonumber \\
\frac{\partial q_{z_{*}}}{\partial k_{x,y}}&=&\frac{\partial\,}{\partial k_{x,y}}
 \left(\,\frac{  q_{\tilde{z}} - k_{\tilde{y}} }{\sqrt{2}}\,\right)  =   -\,
\frac{ k_{\tilde{y}}+ q_{\tilde{z}} }{q_{\tilde{z}}\,\sqrt{2}}\,\left[\,\cos\theta\,
 \frac{\partial k_y}{\partial k_{x,y}} + \sin\theta\,\frac{\partial k_z}{\partial
k_{x,y}} \,\right]
\,,\\
\frac{\partial k_{\tilde{z}}}{\partial k_{x,y}} & = & \left[\,-\,\sin\theta\,
\frac{\partial k_y}{\partial k_{x,y}} + \cos\theta\,\frac{\partial k_z}{\partial
k_{x,y}} \,\right]
\,,\nonumber
\end{eqnarray}
we immediately find
\begin{eqnarray}
\label{phasex}
x & = & -\, \left[\,
\frac{\partial\phi}{\partial k_x}\,\right]_{_{(0,0)}}\hspace*{-0.2cm}=\,0
\end{eqnarray}
and
\begin{eqnarray}
\label{phasey}
y & = & -\, \left[\,
\frac{\partial\phi}{\partial k_y}\,\right]_{_{(0,0)}}\nonumber\\
 & = & b\,\left(\frac{\sin\theta}{\sqrt{n^2-\sin^{2}\theta}}+1\right)\,\cos\theta \,+\,
  c\,\frac{\sin\theta}{\sqrt{n^2-\sin^{2}\theta}}\,\cos\theta\, -\, c\,\sin\theta
\nonumber \\
   & = &
\left[\,b+(b+c)\,\frac{\sin\theta}{\sqrt{n^2-\sin^{2}\theta}}\,\right]\,\cos\theta\,
-\,c\,\sin\theta\,.
\end{eqnarray}
Thus, the SPM, recovering the result (\ref{eq:eqdsl0}),  represents an alternative
way to obtain the geometrical path of optical beams. For partial internal
reflection, see Fig.~1\,(b), the phase (\ref{phase}) is the only phase which contributes to the SPM
calculation, so this shift in the $y$-coordinate is the real shift seen in an
optical experiment. As we shall discuss in the next subsection an {\em additional phase}
appears for total internal reflection ($\theta>\theta_c$) and an {\em additional
shift}, which cannot be predicted by Snell's law, has to be considered.

\subsection*{\small IV.B TOTAL INTERNAL REFLECTION}

As anticipated in the previous section, for $\theta>\theta_c$ an additional phase
comes from the double internal reflection coefficient
\begin{equation}
\left(\frac{q_{z_{_*}}-\,k_{z_{*}}}{q_{z_{_*}}+\, k_{z_{*}}}\right)^2\,.
\end{equation}
Indeed, by observing that
\begin{eqnarray}
k_{z_{*}}^2
&=& \left[\,1-\left(\frac{\sin\theta +
\sqrt{n^2-\sin^{2}\theta}}{\sqrt{2}}\,\right)^{2}\right]\,k^2 +
\mbox{O}\left[k_x^2,k_y\right] \nonumber \\
&=& \left(1 - \frac{n^{2}}{2} - \sin\theta\,\sqrt{n^2-\sin^{2}\theta}\,
\right)\,k^2 + \,\mbox{O}\left[k_x^2,k_y\right]\,,
\end{eqnarray}
 and  using the constraint (\ref{eq:cri1}),  we have  $k_{z_{*}}^2<0$. Consequently,
the {\em additional} phase
\begin{equation}
\label{adphase}
\varphi =  -\,4\,\arctan\left[\,\frac{|k_{z_{*}}|}{q_{z_{_*}}}\,\right]
\end{equation}
has to be included in our calculation. In order to calculate this new contribution,
we start from
\begin{equation}
\frac{\partial \varphi}{\partial {k_{x,y}}} \,= -\,4\, \,
\frac{q_{z_*}^2}{q_{z_*}^2+\,|k_{z_{*}}|^2}\,\frac{\partial}{\partial {k_{x,y}}}
\left[\,\frac{|k_{z_{*}}|}{q_{z_{*}}}\,\right] = -\,
\frac{4}{q_{z_*}^2+\,|k_{z_{*}}|^2}\,\left(\,q_{z_*}\frac{\partial
|k_{z_{*}}|}{\partial {k_{x,y}}} - |k_{z_{*}}| \frac{\partial q_{z_{*}}}{\partial
{k_{x,y}}}\,\right)
\end{equation}
which, by using the relation
\begin{equation*} q_{z_*}^2+\,|k_{z_{*}}|^2 = (n^{2}
-1)k^2\,\Rightarrow\,\frac{\partial |k_{z_{*}}|}{\partial {k_{x,y}}} =
-\,\frac{q_{z_*}}{|k_{z_{*}}|}\,\frac{\partial q_{z_{*}}}{\partial {k_{x,y}}}\,,
\end{equation*}
becomes
\begin{equation}
\label{adphase2}
\frac{\partial \varphi}{\partial {k_{x,y}}} \,=
\frac{4}{|k_{z_{*}}|}\,\frac{\partial q_{z_{*}}}{\partial {k_{x,y}}}\,.
\end{equation}
Finally, the additional shift in the $y$-axis, also known as GH-shift, is
\begin{eqnarray}
\delta & = & -\,
\left[\,\frac{\partial \varphi}{\partial {k_{y}}}\,\right]_{_{(0,0)}}\nonumber \\  &
= &
\frac{4\,\cos\theta\,\left(\,\sin\theta + \sqrt{n^2 - \sin^{2} \theta}\,\right)
}{k\,\sqrt{\left(\,
n^2-2\,+2\,\sin\theta\,\sqrt{n^2-\sin^{2}\theta}\,\right)\,\left(\,n^2 - \sin^{2}
\theta\,\right)}}\,.
\label{GHs}
\end{eqnarray}
The SPM allows one to obtain the $y$-shift of the outgoing beam both for  partial
and total internal reflection,
\begin{equation}
\Delta y =
\begin{cases}
d &  \mbox{for}\,\theta < \theta_c \, \mbox{[\,partial internal reflection\,]\,,}\\
d \,+\, \delta  &   \mbox{for}\,\theta > \theta_c \, \mbox{[\,total internal
reflection\,]\,.}
\end{cases}
\end{equation}
In concluding this section, we recall that the GH shift ($\delta$) cannot be
obtained by using Snell's law. This additional shift is due to the presence of
evanescent waves in the air zones close to the down and up interfaces and cannot be
predicted from geometrical laws\cite{FTIR2003}. It is similar  to the quantum delay
time discussed in quantum mechanics\cite{Delay1983,Delay1992,Delay1997}.

\section*{\normalsize V. CONCLUSIONS AND OUTLOOKS}

In this article, we have shown the value of the SPM as a mathematical tool in
determining  the path of optical beams. Our analysis, which is done in section III
for s-polarized waves, can be immediately extended to p-polarized waves\cite{Wolf1999,Use2011a,Use2011b,Pol2009}
\begin{equation}
T_{_{out}}^{^{(p)}} =  \frac{4 \,n^{2}\,k_{\tilde{z}}
\,q_{\tilde{z}}}{(q_{\tilde{z}} \,+\, n^2\, k_{\tilde{z}})^2} \left(\frac{q_{z_{*}}
\,-\,n^{2}\, k_{z_{*}}}{q_{z_{*}} \,+ \,n^2\, k_{z_{*}}}\right)^{2} \exp[\,i
\,q_{z_{*}}\,b\,\sqrt{2}\, + \,(\,q_{\tilde{z}} \,-\, k_{\tilde{z}}\,)\, c\,]\,.
\end{equation}

For partial internal reflections, the SPM
reproduces the geometrical result predicted by Snell's law, i.e. $\Delta y=d$. For
total internal reflections, the SPM takes into account the GH shift predicting
$\Delta y=d+\delta$. This additional shift is proportional to the wavelength of the
incoming beam, see Eq.\,(\ref{GHs}). The order of magnitude of the GH shift for a
double  total internal reflections is thus relatively small  and this makes any
experimental observations difficult. Note that red ($\lambda \approx .633\,
\mu\mbox{m}$) lasers, whose beam waist is $w_{0}=1$ mm, undergo a shift
$\delta \approx 10^{^{-4}}w_{0}$.  Due to the fact that this shift depends on
the number of internal reflections, it is clear that, to make this experimental
measurement possible\cite{Exp1960}, we have to  amplify this effect by considering,
for example, a band of $N$ dielectric blocks. In this case, the final GH shift will
be given by $N\delta$.

To guarantee two internal reflections in each block, we impose that the
$z_*$-component of the exit point, $P_{_{\rm right}}$, be the same as the incoming
one,  $P_{_{\rm left}}$. This implies, see Eq.(\ref{eq:pright}),
\begin{equation*} c\,\tan[\,\mbox{$\frac{\pi}{4}$}-\psi\,]\,=\,b \end{equation*}
which after simple algebraic manipulations leads to
\begin{equation}
c \,=\, \displaystyle{\frac{n^{2} + 2 \,\sin\theta \,\sqrt{n^2 -
\sin^{2}\theta}}{n^{2} - 2 \sin^{2}\theta} \,b}\,.
\end{equation}

In a forthcoming  work, we intend to analytically solve the integral (\ref{gout})
and obtain the outgoing beam profile. This analytical study can be done by
approximating the transmission coefficient in view of the result obtained in section
IV.

\vspace*{0.8cm}

\noindent
{\small \bf  ACKNOWLEDGEMENTS.}
The authors gratefully thanks the referees  and the editor for their constructive
comments and useful suggestions. The authors also thanks the CNPq (SdL) and Fapesp (SaC)
for the financial support.

\newpage

\begin{figure}
\vspace*{-2cm} \hspace*{-1.3cm}
\includegraphics[width=19cm, height=25cm, angle=0]{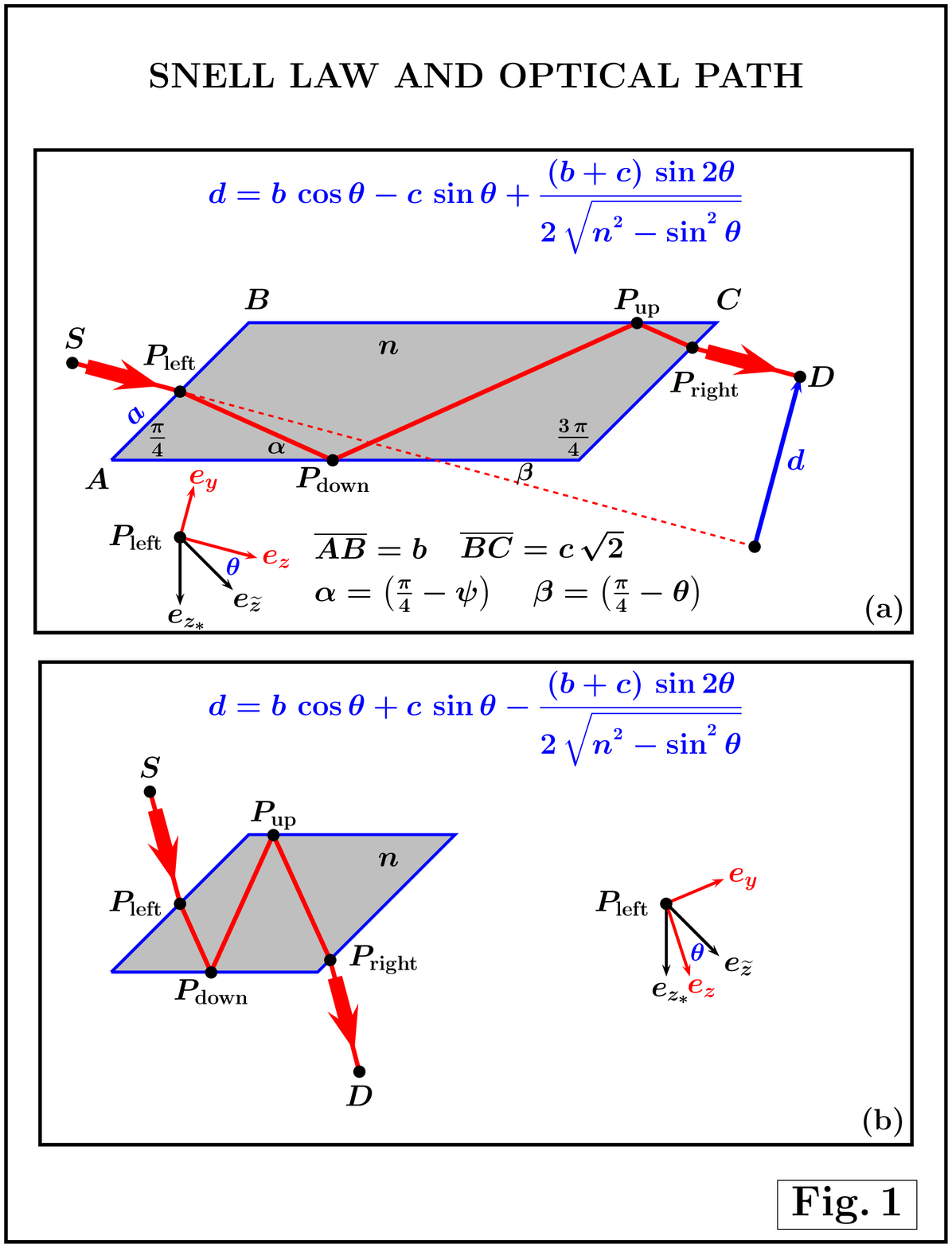}
\vspace*{-2.5cm}
 \caption{Geometric layout of the dielectric block used to determine the optical path by the Snell law. The $\widetilde{z}$ and $z_*$ axes represent, respectively, the normal to the left/right and up/down interfaces. The origin is chosen  at the point in which the incoming beam touches the first interface, $P_{_{left}}$. In (a), the $\widetilde{z}$ axis is obtained from the $z$ axis by a clockwise rotation of angle $\theta$. In (b), the rotation is anticlockwise ($\theta \to -\,\theta$).}
\end{figure}

\newpage
\begin{figure}
\vspace*{-2cm} \hspace*{-1.3cm}
\includegraphics[width=19cm, height=25cm, angle=0]{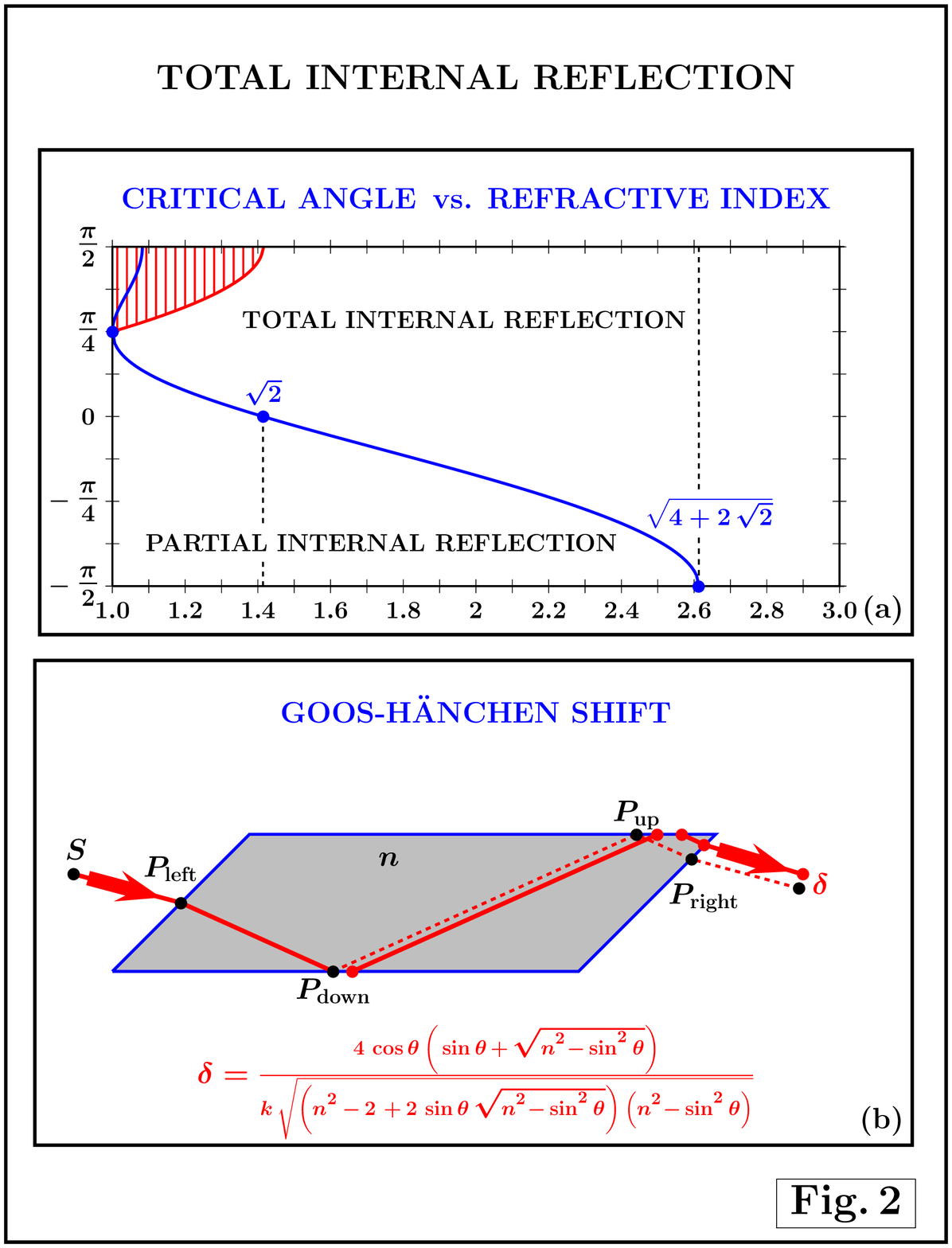}
\vspace*{-2.5cm}
 \caption{In (a), the critical angle, $\theta_c$, is plotted as a function of the refractive index, $n$. The forbidden region represents incidence angles for which the refracted beam at the left interface cannot reach the down boundary. For $n>\sqrt{4+2\,\sqrt{2}}$, we always find total internal reflection. In (b), we can see that, for $\theta>\theta_c$, the geometrical path predicted by the Snell law suffers an additional shift, $\delta$. This additional shift,  known as Goos-H\"anchen shift, can be calculated by using the stationary phase method.}
\end{figure}


\begin{thebibliography}{100}

\bibitem{Wolf1999}
M. Born and E. Wolf,
{\sl Principles of optics} (Cambridge UP, Cambridge, 1999).

\bibitem{Saleh2007}
B.\,E.\,A. Saleh and M.\,C. Teich,
{\sl Fundamentals of Photonics} (Wiley \& Sons, New Jersey, 2007).

\bibitem{Snell2012}
M. Kryjevskaia, M.\,R. Stetzer, and P.\,R.\,L. Heron,
``Student understanding of wave behavior at a boundary: the relationships among
wavelength, propagation speed, and frequency",
Am.\,J.\,Phys. {\bf 80}, 339-347 (2012).

\bibitem{Snell1948}
R. Heller,
``On the teaching of the Snell-Descartes law of refraction",
Am.\,J.\,Phys. {\bf 16}, 356-357 (1948).

\bibitem{Snell1951}
J.\,W. Shirley,
``An Early Experimental Determination of Snell's Law",
Am.\,J.\,Phys. {\bf 19}, 507-508 (1951).

\bibitem{Snell1958}
C.\,V. Bertsch and B.\,A. Greenbaum,
``New apparatus for Snell's law"
Am.\,J.\,Phys. {\bf 26}, 340 (1958).

\bibitem{Snell1980}
H.\,E. Bates,
``An analogue model for teaching reflection and refraction of waves",
Am.\,J.\,Phys. {\bf 48}, 275-277 (1980).

\bibitem{Snell2005}
D. Drosdoff and A. Widom,
``Snell's law from an elementary particle viewpoint",
Am.\,J.\,Phys. {\bf 73}, 973-975 (2005).

\bibitem{Snell2007}
J.\,J. Lynch,
``Snell's law with large blocks",
Phys.\,Teach. {\bf 45}, 180-182 (2007).

\bibitem{Shiff1955}
L.\,I. Shiff,
{\sl Quantum Mechanics} (McGraw-Hill, New York, 1955).

\bibitem{Cohen1977}
C. Cohen-Tannoudji, B. Diu, F. Lalo\"e,
{\sl Quantum Mechanics} (Wiley, Paris, 1977).

\bibitem{RW2009}
S. Longhi,
``Quantum-optical analogies using photonic structures",
Las.\,Phot.\,Rev. {\bf 3}, 243-261 (2009).

\bibitem{RW2013}
X.\,Chen, X.\,J. Lu, Y. Ban, and C.\,F. Li,
``Electronic analogy of the Goos-H\"anchen effect: a review",
J.\,Opt. {\bf 15}, 033001-12 (2013).

\bibitem{QM1957}
L.\,I. Schiff,
``Optical analog of quantum-mechanical barrier penetration",
Am.\,J.\,Phys. {\bf 25}, 207 (1957).

\bibitem{QM2011}
P.\,L. Garrido, S. Goldstein, J. Lukkarinen, and R. Tumulka,
``Paradoxical reflection in quantum mechanics",
Am.\,J.\,Phys. {\bf 79}, 1218-1231 (2011).

\bibitem{QM2012}
G. Zhu and C. Singh,
``Surveying students understanding of quantum mechanics in one spatial dimension",
Am.\,J.\,Phys. {\bf 80}, 252-259 (2012).


\bibitem{Use2008}
S. De Leo and P. Rotelli,
``Localized beams and dielectric barriers",
J.\,Opt. A {\bf 10}, 115001-5 (2008).


\bibitem{Use2011a}
S. De Leo and P. Rotelli,
``Laser interacting with a dielectric block",
Eur.\,Phys.\,J. D {\bf 61}, 481-488 (2011).

\bibitem{Use2011b}
S. De Leo and P. Rotelli,
``Resonant laser tunneling",
Eur.\,Phys.\,J. D {\bf 65}, 563-570 (2011).

\bibitem{Use2013}
M. Selmke and F. Cichos,
``Photonic Rutherford scattering: a classical and quantum mechanical analogy in ray
and wave optics",
Am.\,J.\,Phys. {\bf 81}, 405-413 (2013).

\bibitem{SPM1955}
E. Wigner,
``Lower limit for the energy derivative of the scattering phase shift",
Phys.\,Rev. {\bf 98}, 145-147 (1955).

\bibitem{SPM1973}
R.\,B. Dingle,
{\it Asymptotic expansions: their derivation and interpretation},
(Academic Press, London 1973).

\bibitem{SPM1975}
N. Bleistein and R. Handelsman,
{\it Asymptotic expansions of integrals}, (Dover, New York, 1975).

\bibitem{TIR1981}
D.\,C. Look,
``Novel demonstration of total internal reflection",
Am.\,J.\,Phys. {\bf 49}, 794 (1981).

\bibitem{TIR2005}
 E. Richard V. Keuren,
``Refractive index measurement using total internal reflection",
Am.\,J.\,Phys. {\bf 73}, 611-615 (2005).

\bibitem{GHS1947}
F. Goos and H. H\"anchen,
``Ein neuer und fundamentaler Versuch zur totalreflexion",
Ann.\,der\,Physik {\bf 436}, 333-346 (1947).

\bibitem{GHS1988}
S.\,R. Seshadri,
 ``Goos-H\"anchen beam shift at total internal reflection",
J.\,Opt.\,Soc.\,Am. A \textbf{5}, 583-585 (1988).

\bibitem{GHS2012}
A. Aiello,
``Goos-H\"anchen and Imbert-Federov shifts: a novel perspective",
New\,J.\,of\,Phys. \textbf{14}, 013058-12 (2012).

\bibitem{GHS2013}
K.\,Y. Bliokh and A. Aiello,
``Goos-H\"anchen and Imbert-Fedorov beam shifts: an overview",
J.\,of\,Opt. {\bf 15}, 014001-16 (2013).




\bibitem{WP2008}
M. Andrews,
``The evolution of free wave packets",
Am.\,J.\,Phys. {\bf 76}, 1102-1107 (2008).


\bibitem{Cons1951}
F. Mooney,
``Snell's law equivalent to the conservation of tangential momentum",
Am.\,J.\,Phys. {\bf 19}, 385 (1951).


\bibitem{FTIR2003}
F.\,P. Zanella, D.\,V. Magalhães, M.\,M. Oliveira, R.\,F. Bianchi and L. Misoguti,
``Frustrated total internal reflection: A simple application and demonstration",
Am.\,J.\,Phys. {\bf 71}, 494-496 (2003).


\bibitem{Delay1983}
K. Yasumoto and Y. Oishi,
``A new evaluation of the Goos-H\"anchen shift and associated time delay",
J.\,Appl.\,Phys. \textbf{54}, 2170-2176 (1983).

\bibitem{Delay1992}
W. van Dijk and K.\,A. Kiers,
``Time delay in simple one dimensional systems",
Am.\,J.\,Phys. \textbf{60}, 520-527 (1992).

\bibitem{Delay1997}
L. de la Torre,
``Wave packet distortion and time delay",
Am.\,J.\,Phys. \textbf{65}, 123-125 (1997).


\bibitem{Pol2009}
C. Bahrim and W.\,T. Hsu,
``Precise measurements of the refractive indices for dielectrics using an improved
Brewster angle method",
Am.\,J.\,Phys. {\bf 77}, 337-344 (2009).

\bibitem{Exp1960} N.\,J. Harrick,
``Study of physics and chemistry of surfaces from frustrated total internal
reflections",
Phys.\,Rev.\,Lett. {\bf 4}, 224-226 (1960).


\end{thebibliography}
\end{document}